\documentclass[11pt,letterpaper]{article}
\usepackage[dvips,colorlinks]{hyperref}
\hypersetup{pdftitle={Atomic Fountain},
pdfauthor={Alain Michaud <Alain.Michaud@nrc-cnrc.gc.ca>},
pdfcreator={Latex hyperref},
pdfkeywords={Atomic Clock, Atomic Fountain, Laser Cooling},
bookmarks=true}
\usepackage[dvips]{graphicx}
\title{Realisation of a Frequency Standard \\
        Using an Atomic Fountain\footnote{\normalsize Proceedings of the 7th European Frequency and Time Forum, Neuch\^{a}tel, Switzerland, 16-18 March 1993, pp. 525-530 }}
\author{A. Michaud, M. Chowdhury, K.P. Zetie,  C.J. Cooper \and
 G. Hillenbrand, V. Lorent, A. Steane, and C.J. Foot\footnote{\normalsize University of Oxford, Clarendon Laboratory, 
Oxford, 0X1 3PU, UK }
}
\date{}

\begin{document}

\maketitle
\pdfbookmark[1]{ABSTRACT}{ABSTRACT}

\section*{\label{section:abstract}ABSTRACT}

We report the realisation and preliminary study of a frequency standard using a fountain of laser cooled c\ae sium atoms. Our apparatus uses a magneto-optical trap as a source of cold atoms and optical pumping to prepare the atoms in the correct state before they enter the microwave cavity. \\

\textbf{KEYWORDS}: Atomic clock, atomic fountain, laser cooling.

\section{\label{section:introduction}INTRODUCTION}

C\ae sium atoms at a temperature of only 3$\mu$K can now be produced by laser cooling. These atoms have a velocity of only about 10 mms$^{-1}$ -- a few times the recoil velocity from a single photon. Such ultracold atoms allow much longer measurement times than have previously been possible, together with smaller systematic shifts. These features have been demonstrated for RF and microwave transitions by Kasevich \emph{et al.} \cite{R1} 
\href{http://prola.aps.org/abstract/PRL/v63/i6/p612_1}{[online]}
who made the first working atomic fountain using sodium, and Clairon \emph{et al.} \cite{R2} \cite{R3} \cite{R4} who made a c\ae sium atomic fountain. The advantages of this technique were first considered by Zacharias who attempted to use velocity selection to implement the Ramsey separated oscillatory field method in a fountain by allowing the atoms to interact twice with the same cavity, once on the way up and again on the way down \cite{R5}. This symmetric configuration reduces the cavity phase shift which is one of the most troublesome problems in present primary frequency standards. In this paper, we report on our work to make a frequency standard following this approach. Firstly we give an overall view of the operation of the fountain and then, in the following sections, a more detailled description of those aspects which are specific to our system.    \\

\section{\label{section:overview}OVERVIEW OF EXPERIMENT}

\begin{figure}
\pdfbookmark[1]{Fig. 1: Experimental Arrangement}{fg:setup}
\begin{center}
\includegraphics[width=4in]{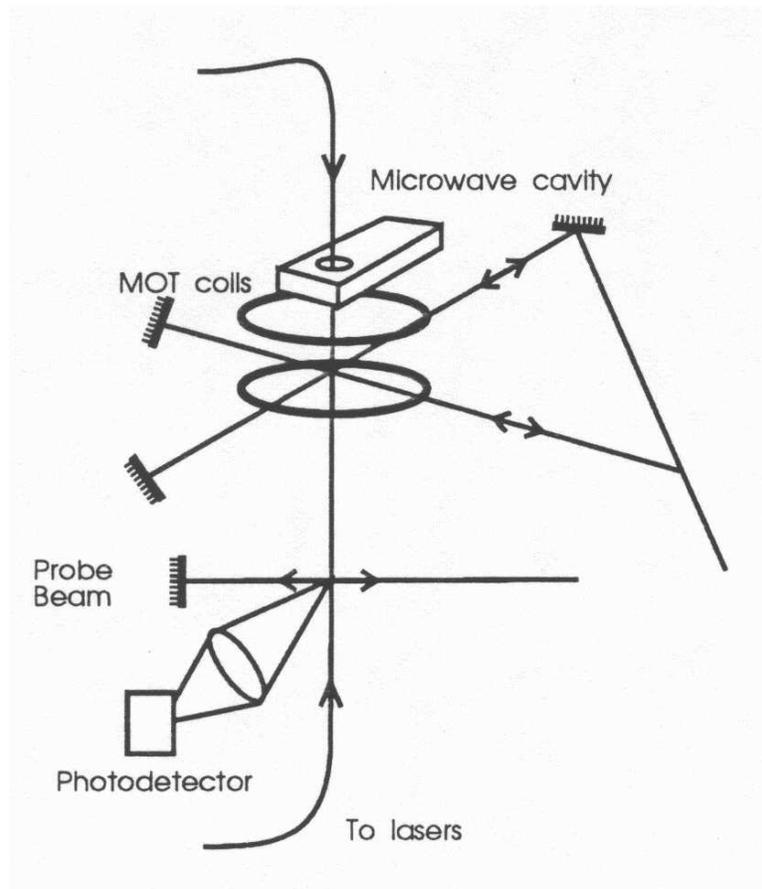}
\end{center}
\caption{\label{fig:setup}Experimental arrangement of the atomic fountain}
\end{figure}

The physical arrangement of the apparatus is shown in figure \ref{fig:setup}. The magneto-optical trap (MOT), used as a source of cold atoms, comprises three orthogonal pairs of laser beams of the requisite polarisations ($\sigma^{+}\ \sigma^{-}$) which intersect in a region where a gradient of magnetic field is produced by a pair of anti-Helmholtz coils \cite{R6} 
\href{http://prola.aps.org/abstract/PRL/v59/i23/p2631_1}{[online]}. Typically there are $10^{7}$ atoms in the trap after a 1 s loading time. These atoms are launched upwards to form a fountain of cold atoms using the following sequence: The magnetic field of the trap is switched off, to leave the atoms in optical molasses. The atoms are then launched upwards by introducing a frequency difference between the vertical beams so that the rest frame of the molasses is moving upwards at the required velocity ("moving molasses technique", \cite{R4}) and the cooling gives a narrow spread in velocities (the laser detuning below resonance is changed smoothly from -10 MHz to -65 MHz during this cooling period). The final stage of the launch sequence is to optically pump all the atoms into the lower hyperfine level (we discuss state selection in section \ref{section:pumping}). After launching, the atoms pass upwards though a 12 mm hole in a microwave cavity (a shorted section of X-band waveguide located 40 mm above the trap centre) and after a period of 0.5 s (for our highest launches) they fall back under gravity passing through the cavity a second time. Both the atoms which have undergone a transition and the total number of atoms are measured by monitoring the fluorescence from a 6 mm diameter probe beam (the photodiode and lens have a combined efficiency of 5\%).\\

\section{\label{section:pumping}OPTICAL PUMPING AND STATE SELECTION}

\begin{figure}
\pdfbookmark[1]{Fig. 2: Energy Levels}{fg:levels}
\begin{center}
\includegraphics[width=4in]{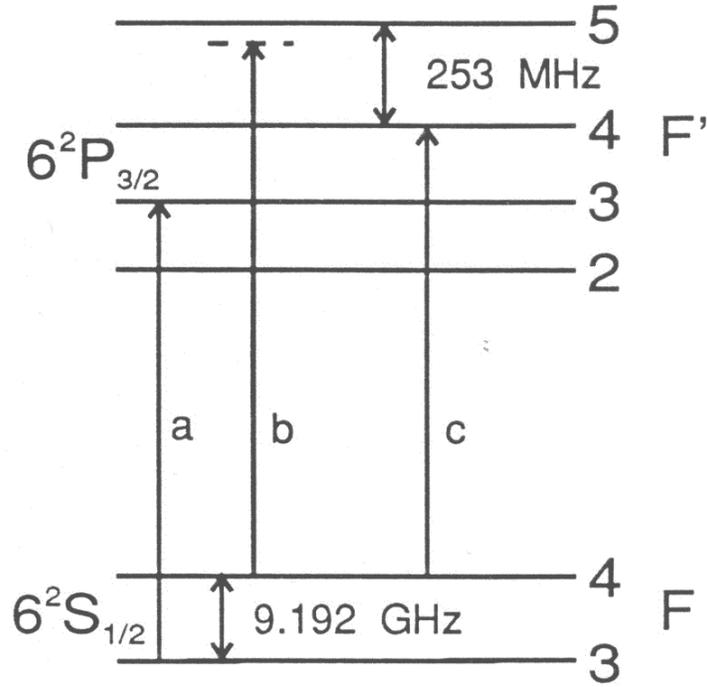}
\end{center}
\caption{\label{fig:levels}Energy level scheme $D_{2}$ line in c\ae sium. The transitions used are: (a) repumper, (b) trapping and cooling, (c) pumping.}
\end{figure}

During the loading of the trap and launching by molasses, two laser beams are present. The main beam is tuned slightly below the F=4 $\rightarrow$ F'=5 cycling transition (figure \ref{fig:levels}) and provides the cooling and trapping forces. Off-resonnant excitation of the F=4 $\rightarrow$ F'=4 transition causes some of the atoms to leak into the lower hyperfine level (F=3), where they are lost. Therefore, a second laser beam is necessary, tuned to the F=3 $\rightarrow$ F'=3 or F'=4 transitions, to repump those atoms.\\

C\ae sium atomic clocks are based on the magnetic dipole transition $\left| 3,0 \right>$ to $\left| 4,0 \right>$ between the ground state hyperfine levels since neither of these states has any first order Zeeman shift \cite{R7}. Thus to observe the clock transition there must be a state selection process to create a population imbalance between these two states. The current working primary time standards use magnetic state selection but optical pumping is the the natural choice for laser cooled atoms since no additional lasers are required. The advantages of optical pumping have long been recognized and there are a considerable number of published papers in which various techniques have been analyzed theoritically and experimentally \cite{R8}
\href{http://www.iop.org/EJ/abstract/0026-1394/29/2/003}{[online]}. Optical pumping methods for clocks can be divided into two categories, namely one-laser and two-laser schemes; we have used both types.\\

In a simple one-laser scheme all the atoms are transfered into one of the hyperfine levels. This hyperfine-pumping was carried out in our initial experiments (and those of Clairon \emph{et al.}) by using a laser resonant with the F=4 $\rightarrow$ F'=4 transition which emptied the upper level into F=3 with each atom scattering an average of 2.4 photons. We have achieved effective pumping by this method by simply cutting the repumping beam after the cooling period in molasses (when the main laser is detuned 65 MHz below F=4 $\rightarrow$ F'=5) and deliberately allowing the atoms to leak out of the cooling cycle. Experimentally we have found that this alternative method does not always give complete pumping and the slower rate of pumping will lead to a spread in the vertical velocity distribution (this is possibly an advantage since it reduces the atomic density without atoms being lost by transverse spreading). However this simplified method can be used as part of a two-laser scheme.\\

Two-laser schemes exploit the fact that the electric dipole matrix element between two $M_{F}$ = 0 states of the same total angular momentum is zero e.g. $\left< F = 3, M_{F} =0 \mid \mathbf{ d   \cdot e }\mid F' = 3, M_{F} =0 \right> = 0$ for the scheme which we consider here. Thus the combisation of one laser effecting hyperfine-pumping out of the F=4 level and a second laser resonant with the $F=3 \rightarrow F'=3$ transition (linearly polarized along the B-field) wich excites $\pi$-transitions, the atoms cycle many times until they fall into the only state not being excited, $\left| F = 3, M_{F} =0 \right>$. Although a high degree of pumping has been achieved in thermal beams \cite{R8} 
\href{http://www.iop.org/EJ/abstract/0026-1394/29/2/003}{[online]}
 it involves the scattering of many photons, and would cause considerable heating. However a compromise can be achieved by only allowing a few optical pumping cycles since a worthwile number of atoms are transfered into $M_{F}=0$ after scattering up to two photons and those already in this state are not heated at all. Other possible schemes are discussed in \cite{R8}
\href{http://www.iop.org/EJ/abstract/0026-1394/29/2/003}{[online]}, in particular some which use $\pi$-pumping on the F=4 $\rightarrow$  F'=4 transition. The increase in the signal for a given atomic number density will be particularly important if collisional shifts are significant.\\

\section{\label{section:system}OPTICAL SYSTEM}

The requisite laser beams were obtained using only two independant diode laser systems. Both systems were based on high power Spectra Diode lasers frequency stabilised by an external grating cavity \cite{R9} 
\href{http://dx.doi.org/10.1016/0030-4018(85)90041-0}{[online]},
with reference to the saturated absorbtion spectrum of c\ae sium. The output in zeroth order of the grating was about 10 mW, which was sufficient for the repumping beams. To obtain more power for the main cooling beams the light was amplified to 100 mW by injection locking another diode, and a further increase in power could realily be achieved by using the same master to control several slave lasers whilst still maintaining the long term reliability required for a clock. The master laser was computer controled so that its frequency could either be locked at a fixed offset from the F=4 $\rightarrow$ F=5 saturated absorption line or moved in less than one 1 ms to any nearby frequency.\\

\begin{figure}
\pdfbookmark[1]{Fig. 3: Laser Systems}{fg:lasers}
\begin{center}
\includegraphics[width=3.5in]{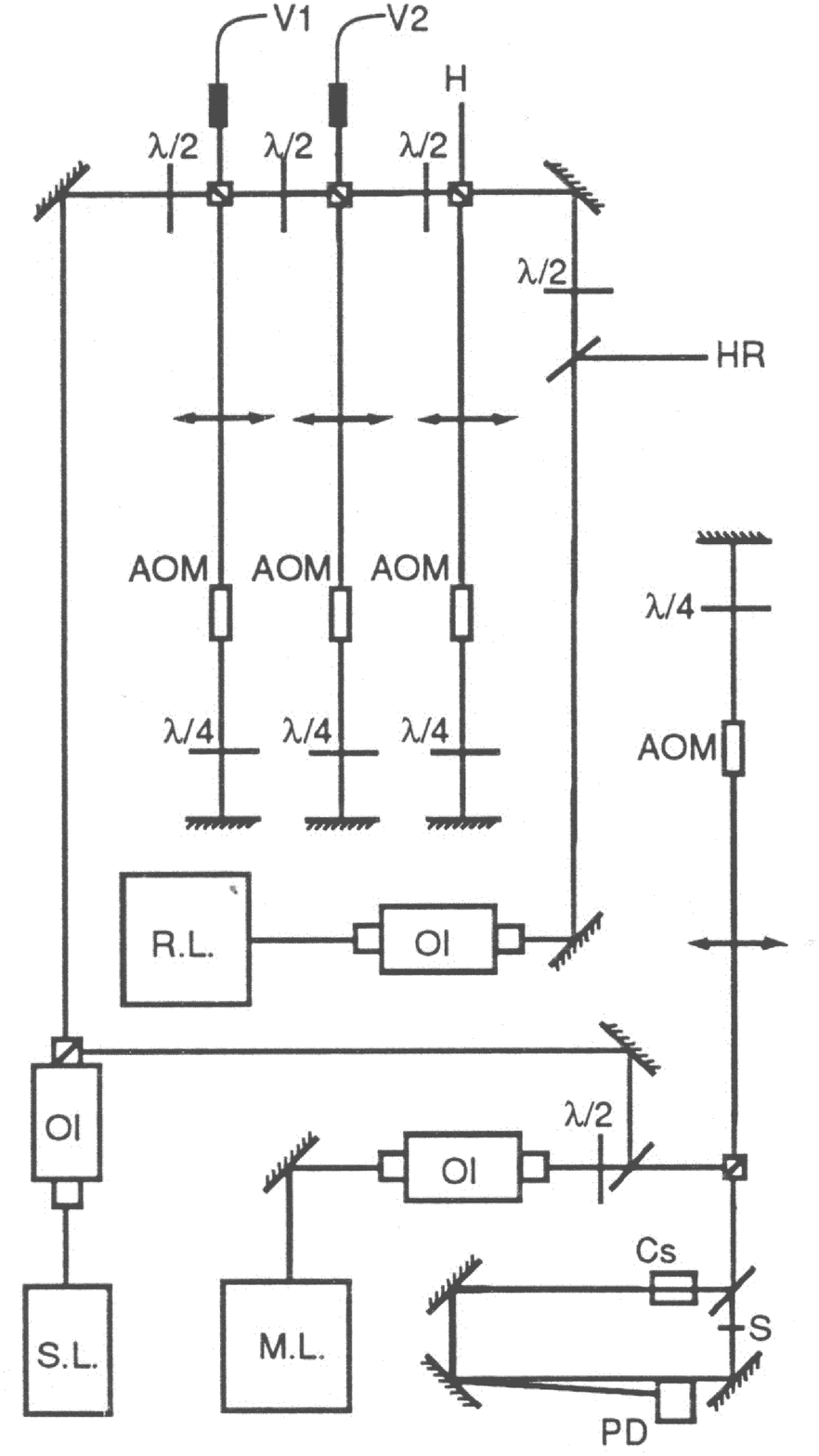}
\end{center}
\caption{\label{fig:lasers}Laser Systems: AOM: Acousto-Optic Modulator, Cs: C\ae sium Cell, ML: Master Laser, P: Photodetector, RL: Repumper Laser, S: Shutter, SL: Slave Laser, OI: Optical Isolator, $\lambda$/2, $\lambda$/4: Retardation plates}
\end{figure}

Figure \ref{fig:lasers} shows the schematic of the lasers and the associated optics. The light from the slave laser is double-passed through acousto-optic modulators to enable a frequency shift to be introduced between vertical beams and to shut off all the beams quickly. The AOMs shift the frequency of the light, relative to the laser frequency, by 2$\times$80 MHz in the horizontal beams (H) or 2 $\times$(80$\pm \delta$) for the vertical beams (V1,V2). In order that these output beams have a small detuning below the required transition we used another AOM at a frequency of 2 $\times$ 75 MHz before the absorption cell.  The frequency of this AOM is dithered to allow phase sensitive detection of the saturated absorption spectrum in the cell. Part of the repumping laser beam is mixed with the horizontal beam (H) and some was transmitted separately (HR) to the clock for $\pi$-polarized beam required for the second scheme described in section \ref{section:pumping}. Optical fibres for the vertical beams eliminate possible movements from the misalignment of the back reflection through the AOMs and ensure good beams quality.\\

\section{\label{section:results}RESULTS}

\begin{figure}
\pdfbookmark[1]{Fig. 4: Ramsey Fringe}{fg:fringe}
\begin{center}
\includegraphics[width=4in]{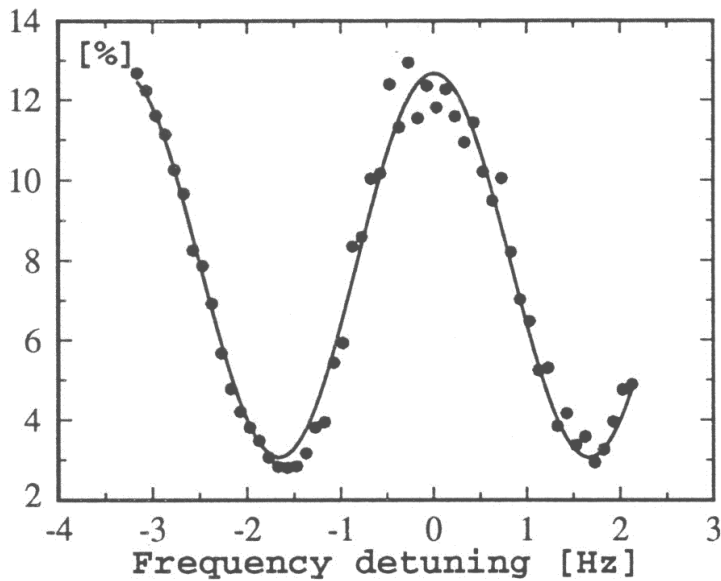}
\end{center}
\caption{\label{fig:fringe}Central Ramsey fringe corresponding to a 11 cm launch above the cavity.}
\end{figure}

\begin{figure}
\pdfbookmark[1]{Fig. 5: 2 Lasers Patern}{fg:pattern}
\begin{center}
\includegraphics[width=4in]{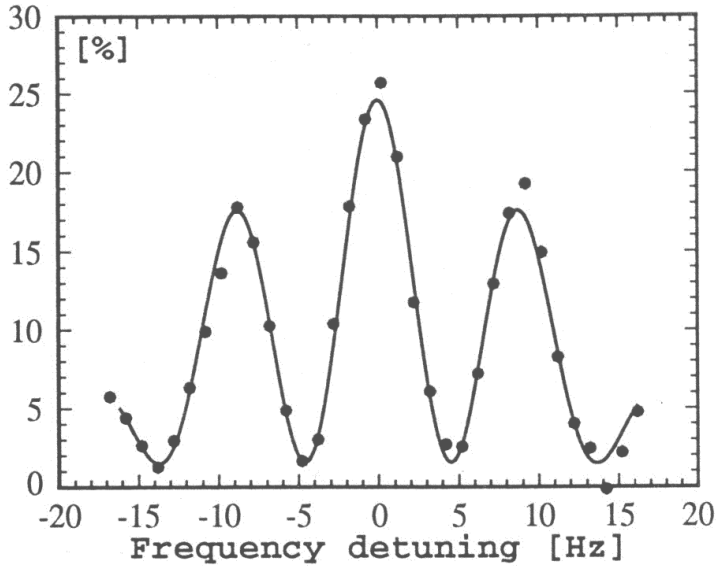}
\end{center}
\caption{\label{fig:pattern}Ramsey pattern obtained with two laser pumping.}
\end{figure}

\begin{figure}
\pdfbookmark[1]{Fig. 6: Amplitude vs Angle}{fg:fringeamplitude}
\begin{center}
\includegraphics[width=4in]{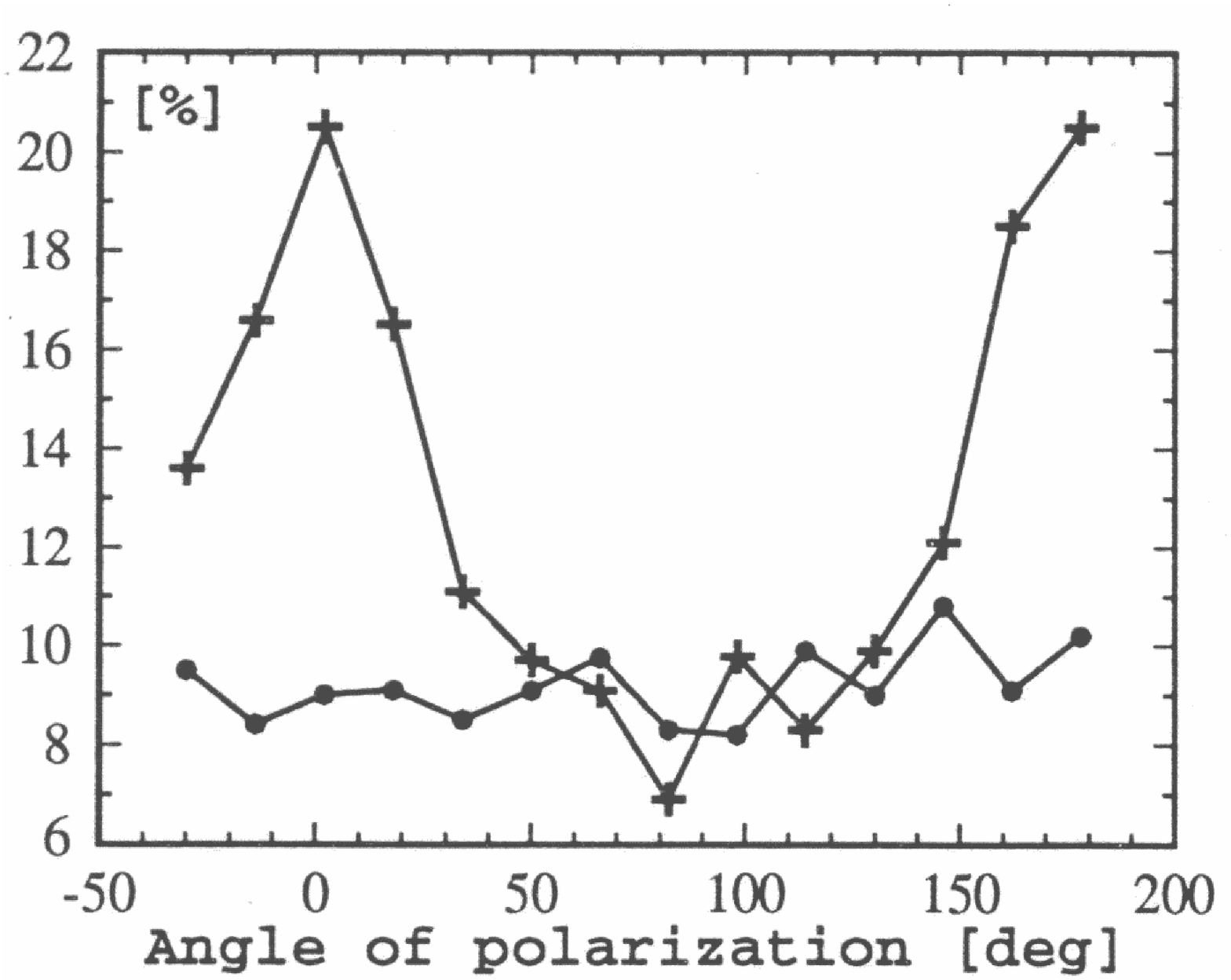}
\end{center}
\caption{\label{fig:fringeamplitude}Fringe amplitude as a function of the angle of polarisation of the repumping light ($+$), and without repumping light ($\bullet$)}
\end{figure}

As an example of the possibilities of this of this new type of frequency standard, we show some Ramsey patterns obtained using this apparatus. The fraction of atoms detected in the level $\left| F = 4, M_{F} =0 \right>$ is plotted as a function of the microwave frequency detuning in Hz. Figure \ref{fig:fringe} corresponds to a launch of 110 mm above the cavity, corresponding to a width of 1.7 Hz, obtained using one-laser pumping (maximum of 14\% of atoms in $M_{F}=0$). Each point is an average of 10 measurement cycles and the S/N is 30. Figure \ref{fig:pattern} corresponds to a launch of 57 mm obtained using the two-laser pumping scheme. To verify the mechanism the signal at the maximum of the central fringe was measured as a function of the angle of polarisation of the repumper relative to the bias B-field (5$\mu$T in the horizontal direction)(figure \ref{fig:fringeamplitude}). This is a measurement of the number of atoms in the $\left| F = 3, M_{F} =0 \right>$ level and shows, that the relative number of atoms in this state can be increased by a factor of $\sim$2.5 without significant heating.\\

\begin{figure}
\pdfbookmark[1]{Fig. 7: Constant Excitation}{fg:leakeage}
\begin{center}
\includegraphics[width=4in]{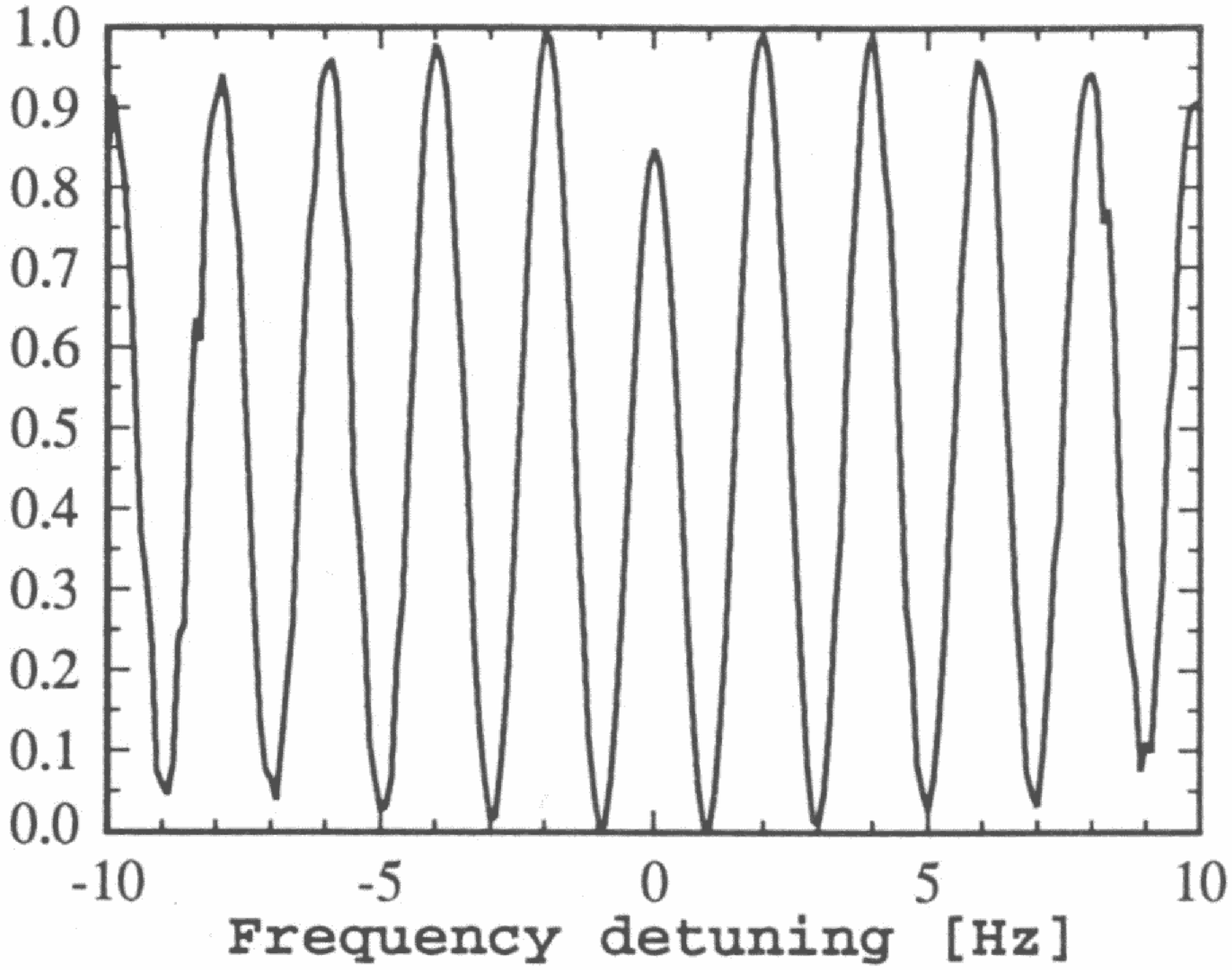}
\end{center}
\caption{\label{fig:leakeage}Ramsey pattern showing the effect of a small, constant microwave excitation.}
\end{figure}

We also observed anomalies on some experimental patterns which cannot be explained using the standard model. We beleive that the central fringe amplitude may be affected by leakage of microwave from the waveguide. In order to illustrate this effect, Figure \ref{fig:leakeage} shows a pattern obtained by numerical simulation. For a typical launch (T-0.5 s, $\tau$=2 ms, $b\tau=\pi$) and the velocity distribution is assumed gaussian: $\sigma$ = 10 ms) the atoms were submitted to an excitation of the amplitude 0.002 $\times  b$ during the period $T$ (we assumed a constant amplitude and phase). We clearly see the collapse of the central fringe. This behavior is different to that due to incorrect power level as it affects mainly the central fringe.\\

This result emphasizes the need for a well designed microwave cavity and further investigation of a possible frequency shift. \cite{R8}
\href{http://www.iop.org/EJ/abstract/0026-1394/29/2/003}{[online]}\\

\section{\label{section:conclusion}CONCLUSION}

We have realised a prototype frequency standard based on an atomic fountain of cold c\ae sium atoms and have obtained Ramsey patterns with linewidth below 2 Hz. A large part of the observed noise is due to effects of fluctuations of time of arrival ($\pm$ 1 \%) and the normalization technique used; this will be reduced by the use of a second probe beam to provide improved normalisation of the signal. We have also shown that the ratio of atoms present in the $\left| F = 3, M_{F} =0 \right>$ level can be increased  without significant heating. Although no systematic investigations have been done, it has been observed that the S/N is also increased in the same proportion.\\

This research is sponsored by the National Physical Laboratory. KPZ and AMS are research fellows at Christ Church and Merton College respectively. We acknowledge valuable discussions with A. Clairon, D. Knight, P. Whibberly and C. Salomon\\

\pdfbookmark[1]{REFERENCES}{REFERENCES}

\bibliographystyle{unsrt}

\bibliography{fountain}

\renewcommand{\refname}{REFERENCES}

\end{document}